\documentclass{eptcs}
\providecommand{\event}{ADG 2021} 
\usepackage{graphicx}
\usepackage{alltt}
\usepackage{url}
\usepackage{listings}

\def\orcidID#1{\unskip$^{\mbox{\href{https://orcid.org/#1}{\scriptsize{[#1]}} }}$}

\title{Spreads and Packings of PG(3,2), Formally!}
\author{Nicolas Magaud\orcidID{0000-0002-9477-4394}
\institute{ICube UMR 7357 CNRS \\
Université de Strasbourg, France}
\email{magaud@unistra.fr}}

\def\titlerunning{Spreads and Packings of PG(3,2), Formally!}
\def\authorrunning{Magaud}

\begin{document}
\maketitle              

\begin{abstract}
  We study how to formalize in the Coq proof assistant the smallest
  projective space PG(3,2). We then describe formally the spreads and
  packings of PG(3,2), as well as some of their properties. The formalization is rather
  straightforward, however as the number of objects at stake increases
  rapidly, we need to exploit some symmetry
arguments as well as smart proof techniques to make proof search and
verification faster and thus tractable using the Coq proof
assistant.  This work can be viewed as a first
  step towards formalizing projective spaces of higher dimension,
  e.g. PG(4,2), or larger order, e.g. PG(3,3). 

\end{abstract}

\section{Introduction}
Projective incidence geometry \cite{coxeter,Buekenhout95} is one of
the simplest description of geometry, where only points and lines as
well as their incidence properties are 
considered.  In addition, in such a setting, we assume that two
coplanar lines always meet.  There exist some finite and infinite
models of projective incidence geometry. Finite projective spaces have been
studied extensively from a mathematical point of view~(see e.g
\cite{HirschfeldJ.W.P.JamesWilliamPeter1985Fpso}). Recently
\cite{BMS18b}, we started studying small finite projective planes/spaces from a computer
science perspective.  We formalized usual projective
planes such as PG(2,2), PG(2,3) or PG(2,5) as well as the smallest projective space PG(3,2) using the Coq
proof assistant~\cite{coqmanual,BC04}. We especially focused on proving the synthetic
axioms for projective geometry hold in these models.  
In this paper, we study some of the characteristic subsets of PG(3,2), such as spreads of lines and
packings of spreads.

In a three-dimensional setting, the number of points and lines increase rapidly
with the order, as shown in
Fig.~\ref{numbers}. Thus we need
to design extremely efficient proof techniques for PG(3,2) if we want
our approach to be scalable to
projective spaces of higher dimension or larger order.
The whole Coq formalization is available online and can be retrieved
at : \url{https://github.com/magaud/PG3q}. Pointers to specific parts
of the development are given throughout the paper.

\nocite{Buekenhout95} \nocite{coxeter}

\begin{figure}
$$
\begin{array}{|c|c|c|c|}
\hline
& \textsf{\# points} & \textsf{\# lines} & \textsf{\# points per line}\\
  PG(2,2)  & 7 & 7 & 3\\
  \hline
  PG(2,3)  & 13 & 13 &4 \\
  \hline
PG(2,5) &  31 & 31 & 6\\
  \hline
  PG(2,n)& n^2+n+1&n^2+n+1&n+1\\
  \hline
  \hline
PG(3,2) & 15 & 35 &3\\
\hline
PG(3,3) & 40 & 130& 4\\
\hline
PG(3,4) & 85 & 357& 5\\
\hline
PG(3,q) & (q^2+1)(q+1) &(q^2+q+1)(q^2+1)& q+1\\
  \hline
\end{array}
$$
\caption{Numbers of points, lines and points per line depending on the
dimension and the order of projective planes and spaces\label{numbers}}
\end{figure}

This paper is organized as follows. In Sect.~\ref{specification}, we show how to formally describe PG(3,2) in Coq using
plain inductive types. In Sect.~\ref{spreads}, we compute all the spreads and
packings of PG(3,2) and prove some of their properties.  In
Sect.~\ref{engineering}, we present some of the proof optimization techniques we set up
in order to achieve the proofs. Finally, in Sect.~\ref{conclusion}, we outline
how this work can be extended to projective spaces of higher dimension
or higher order.  
\section{Inductive specification of PG(3,2) in Coq}\label{specification}

\begin{figure}
\begin{alltt}
Parameter Point, Line : Type.

Parameter eqP : Point -> Point -> bool.
Parameter eqL : Line -> Line -> bool.

Parameter incid_lp : Point -> Line -> bool.
  
Definition Intersect_In (l1 l2 :Line) (P:Point) := 
  incid_lp P l1 && incid_lp P l2. 

Definition dist_3p  (A B C :Point) : bool := 
  (negb (eqP A B)) && (negb (eqP A C)) && (negb (eqP B C)).

Definition dist_4p  (A B C D:Point) : bool :=
     (negb (eqP A B)) && (negb (eqP A C)) && (negb (eqP A D))
  && (negb (eqP B C)) && (negb (eqP B D)) && (negb (eqP C D)).

Definition dist_3l (A B C :Line) : bool :=
  (negb (eqL A B)) && (negb (eqL A C)) && (negb (eqL B C)).

Axiom a1_exists : forall A B : Point,
                    \{l : Line| incid_lp A l && incid_lp B l\}.

Axiom uniqueness : forall (A B :Point)(l1 l2:Line),
      incid_lp A l1 -> incid_lp B l1  ->
      incid_lp A l2 -> incid_lp B l2 -> A = B \/ l1 = l2.

Axiom a3_1 : forall l:Line,
  \{A:Point & \{B:Point & \{C:Point |
    (dist_3p A B C) && (incid_lp A l && incid_lp B l && incid_lp C l)\}\}\}.

Axiom a2 : forall A B C D:Point, forall lAB lCD lAC lBD :Line,
             dist_4p A B C D -> 
             incid_lp A lAB && incid_lp B lAB ->
             incid_lp C lCD && incid_lp D lCD ->
             incid_lp A lAC && incid_lp C lAC ->
             incid_lp B lBD && incid_lp D lBD ->
             (exists I:Point, incid_lp I lAB && incid_lp I lCD) ->
             exists J:Point, incid_lp J lAC && incid_lp J lBD.

Axiom a3_2 : exists l1:Line, exists l2:Line,
               forall p:Point, ~(incid_lp p l1 && incid_lp p l2).

Axiom a3_3 : forall l1 l2 l3:Line,
  dist_3l l1 l2 l3 ->
  exists l4 :Line,  exists J1:Point, exists J2:Point, exists J3:Point,
    Intersect_In l1 l4 J1 && Intersect_In l2 l4 J2 && Intersect_In l3 l4 J3.
\end{alltt}
\caption{Projective spaces of \label{spec} dimension 3: definitions and properties (\texttt{pg3x\_spec.v})}

\end{figure}

\subsection{Definitions and operations}
We choose to use two simple inductive types to represent points and
lines of PG(3,2). Points are represented by an inductive datatype of
15 constructors without arguments. Lines are represented in the same way
using 35 constructors.  As there are three points per line, the incidence
relation can be represented in a compact way using the \texttt{match
  ... with} construct of Coq specification language.
\begin{verbatim}
Inductive Point :=
| P0 | P1 | P2 | P3 | P4 | P5 | P6 | P7 | P8 | P9 
| P10 | P11 | P12 | P13 | P14 .

Inductive Line :=
| L0 | L1 | L2 | L3 | L4 | L5 | L6 | L7 | L8 | L9 
| L10 | L11 | L12 | L13 | L14 | L15 | L16 | L17 | L18 | L19 
| L20 | L21 | L22 | L23 | L24 | L25 | L26 | L27 | L28 | L29 
| L30 | L31 | L32 | L33 | L34 .
\end{verbatim}

\begin{verbatim}
Definition incid_lp (p:Point) (l:Line) : bool := 
match l with 
| L0 => match p with P0 | P1 | P2 => true | _ => false end
| L1 => match p with P0 | P3 | P4 => true | _ => false end
| L2 => match p with P0 | P5 | P6 => true | _ => false end
| L3 => match p with P0 | P7 | P8 => true | _ => false end
| L4 => match p with P0 | P10 | P9 => true | _ => false end
| [...]
end.
\end{verbatim}

In order to avoid writing too much code in Coq, we choose to build a
generator (a simple C program) which takes as input the number of
points, the number of lines as well as the incidence relation as a
plain file (\texttt{pg32.txt}\footnote{\url{https://github.com/magaud/PG3q/blob/master/pg32/pg32.txt}}) which contains for each line of the projective space, the
list of the points which are incident to it.  
Given these three elements, the system
automatically builds the inductive data-type for points and lines, the
incidence relation. It also defines an artificial order on points and lines based on
the index of corresponding points and lines, i.e. $\texttt{P0} < \texttt{P1} < \texttt{P2} <
\ldots < \texttt{P14}$.  The specification generator also builds some
auxiliary functions, which will be useful
to prove existential statements of the form $\forall l1~
l2:\texttt{Line}, ~exists ~P:\texttt{Point}, ~\ldots$.

Using plain inductive data-types may seem naive. An alternative approach to specify points and lines of PG(3,2) could
be to use finite types $'I_n$  of ssreflect and the mathematical components
library \cite{gonthier:inria-00258384,math-components}.
However the main drawback is that ssreflect is designed for formal
reasoning rather than computing.  Thus checking the incidence between a point
and a line is a highly expensive operation, which prevents us from
carrying out proofs efficently. Using plain inductive types is much
more efficient both to check incidence properties and to perform case analysis. The only drawback is that inductive data-types and
functions are huge to write, but this is not that important as we
manage to generate these specifications automatically.
Overall, our choice is to use the main features of ssreflect, especially the
small-scale reflection pattern, but with our own datatypes.

\subsection{Proofs}
Once the projective space PG(3,2) is described, we check whether all
the axioms for projective space geometry hold for this model.
This requires proving all axioms of the module defined in
\url{https://github.com/magaud/PG3q/blob/master/generic/pg3x_spec.v}
and presented in Figure~\ref{spec}.
This is pretty straightforward and we try and make it as generic and
efficient as
possible so that it can be reused for other models of projective space
such as PG(3,3).

\section{Spreads and Packings of PG(3,2)}\label{spreads}
\subsection{Definitions}
A spread of PG(3,q) is a set of $q^2+1$ lines which are pairwise disjoint
and thus partition the set of points. In PG(3,2), it corresponds to some sets
of 5 lines. A packing of PG(3,q) is a set of $q^2+q+1$ spreads which are
pairwise disjoint and thus partition the set of lines. In PG(3,2), it
corresponds to some sets of 7 spreads. 
As recalled in
\cite{DBLP:journals/dcc/Betten16,cole1922,JEURISSEN1995617}, it is
well known that there is only one spread (up to isomorphism) in
PG(3,2) and two packings (up to isomorphism).

\subsection{Generating all Spreads and Packings of PG(3,2)}
Using our external program, we automatically generate all sets of lines of PG(3,2) which are
disjoint and cover all the points. As lines contains exactly 3 points,
they need to be sets of exactly 5 lines so that all the points of
PG(3,2) are accounted for. We obtain 56 distinct spreads
(modulo permutations of the order of the lines involved). We also
generate all sets of spreads which are disjoint and cover all the
lines. As before, these sets of spreads must have 7 elements, as the number
of spreads multiplied by the number of lines in each spread equals the
number of lines (35) of PG(3,2). As expected, see Theorem 17.5.6 in \cite{HirschfeldJ.W.P.JamesWilliamPeter1985Fpso}, we find 240 packings,
upto isomorphism.
Spreads and packings are grouped in lists. The list of spreads contain
56 spreads, each of them being a list of 5 lines. The list of packings
contain 240 packings, each of them being a list of 7 spreads.
\begin{verbatim}
Definition S0 := [ L0; L19; L24; L28; L33 ].
Definition S1 := [ L0; L19; L26; L29; L32 ].
[...]
Definition spreads := [ S0 ; S1 ; S2 ; ... ; S54; S55 ].
\end{verbatim}

\subsection{Properties}

In Coq, we easily check that the computed spreads and packings
verify the properties of spreads and packings. 

Spreads can be specified using the following definitions  \texttt{is\_partition}
and \texttt{is\_spread5}. The function \texttt{forall\_Point} is a finite universal quantification, and
\texttt{forall\_Point (fun t => X t)} stands for $\texttt{X} ~\texttt{P0} ~\&\&~ \texttt{X} ~\texttt{P1} ~\&\&~ \texttt{X}~ \texttt{P2} ~\ldots~ \&\& ~\texttt{X}~ \texttt{P14}$.  
\begin{alltt}
  Definition is_partition (p q r s t: list Point) :bool :=
  (forall_Point
  (fun x => inb x p || inb x q || inb x r || inb x s || inb x t))
  &&
  (forall_Point
  (fun x => negb (inb x p && inb x q && inb x r &&
                            inb x s && inb x t))).

Definition is_spread5 (l1 l2 l3 l4 l5:Line) : bool :=
  disj_5l l1 l2 l3 l4 l5  &&
  is_partition (all_points_of_line l1) (all_points_of_line l2)
                (all_points_of_line l3) (all_points_of_line l4)
                (all_points_of_line l5).
\end{alltt}
Once these definitions are set, we can prove that the spreads of
PG(3,2) are exactly the ones automatically generated by our external
program.  
\begin{verbatim}
Lemma is_spread_descr : forall l1 l2 l3 l4 l5, 
  (is_spread5 l1 l2 l3 l4 l5) <-> In [l1;l2;l3;l4;l5] spreads.
\end{verbatim}
In addition, we can prove that all 56 spreads of PG(3,2) are
isomorphic. It can be expressed by stating that there exists a
collineation, i.e.  an automorphism of PG(3,2) which respects incidence, between any two spreads of PG(3,2). 

\begin{verbatim}
Lemma all_isomorphic_lemma : forall t1 t2 : list Line, 
  In t1 spreads -> In t2 spreads -> are_isomorphic t1 t2. 
\end{verbatim}
              
To prove this statement, we show it is equivalent to simply proving
that there exists a
collineation (we actually build it) from the $n$-th element of the
list to the $(n+1~\texttt{mod}~56)$-th
element of the list\footnote{\url{https://github.com/magaud/PG3q/blob/master/pg32/pg32\_spreads_collineations.v}}.

It is more challenging
to check that there are (up to isomorphism) only two packings.
\section{Proof Engineering Techniques}\label{engineering}
\subsection{Using bool instead of Prop}
As we work with finite types, equality and the other relations that we use
are decidable. We can directly implement such relations as operations producing elements of the boolean datatype \texttt{bool}. This is more
convenient than defining them as operations producing elements of type
\texttt{Prop} together with a decidability property: $\forall~ x~
y,\{x=y\}+\{\lnot x=y\} $. This practical approach is inspired by the
ssreflect~\cite{GM08} and the mathematical components~\cite{math-components} libraries.

In this setting, logical reasoning (eliminating conjunctions or
disjunctions) is a bit more technical. However this makes most proofs much easier to
complete by simply computing a boolean value and checking that it is
equal to \texttt{true}.  
\subsection{Without Loss of Generality}
Most proofs are highly branching. For instance, performing case analysis on all
three lines to prove the lemma \texttt{a3\_3} leads to $35^3 = 42 875$ cases.  In order to make the
proof more tractable, we use a tactic named \texttt{wlog}\footnote{\url{https://github.com/magaud/PG3q/blob/master/generic/wlog.v}}, which implements the
\textsl{without loss of generality} principle, as it is described in
\cite{DBLP:conf/tphol/Harrison09}.

This allows to reduce the number of
cases to solve explicitly. To use it, we build a virtual order on the points and
lines, simply mapping point \texttt{Pi} (resp. line \texttt{Li}) to
the value $i$ of its index and extends our statement of the form
$\forall l1, l2 :\texttt{Line},\ldots$ to $\forall l1, l2
:\texttt{Line},l1 <l2 \rightarrow\ldots$ .  

\begin{figure}[!t]
\begin{center}
  \includegraphics[width=.6\textwidth]{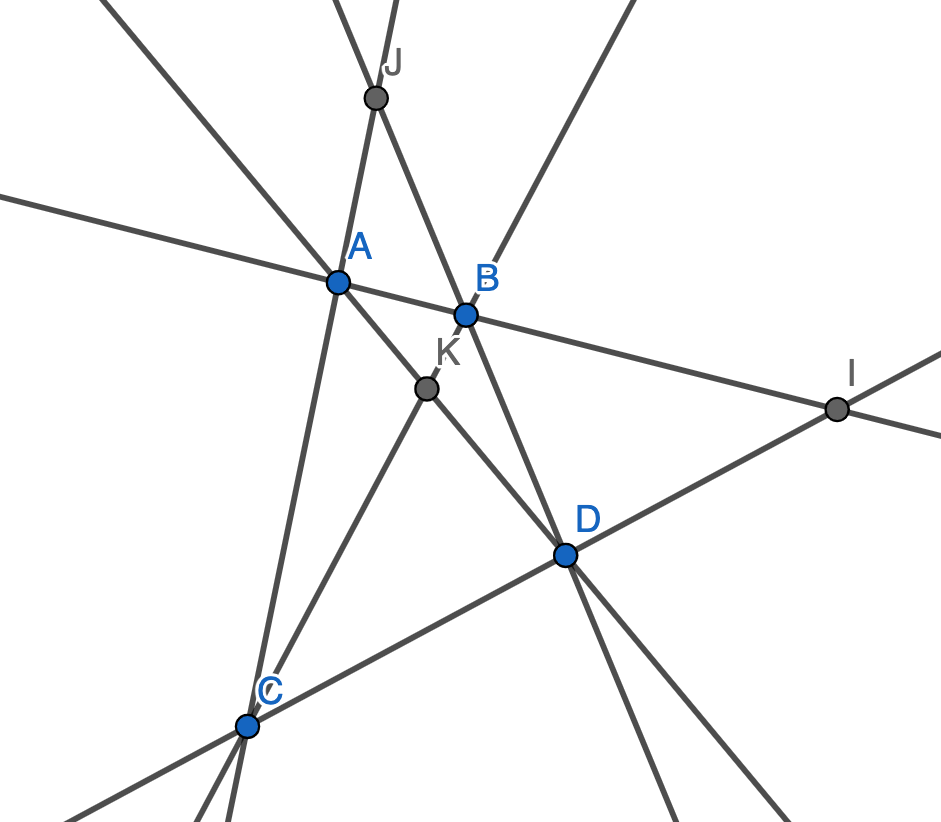}
  \caption{An illustration of the new form of Pasch's axiom used to
    deal with symmetries}\label{Pasch}
  \end{center}
\end{figure}

Surprisingly, using the without loss of generality tactic forces us to
generalize our statement for Pasch's axiom to accommodate all cases,
depending on the order in which we consider points $A$, $B$, $C$, and $D$, as
shown in Fig.~\ref{Pasch}.

The usual conclusion of Pasch's axiom
\begin{alltt}
(exists I:Point, incid\_lp I lAB \&\& incid\_lp I lCD) ->
        {exists J:Point}, incid\_lp J lAC \&\&  incid\_lp J lBD.
      \end{alltt}
      is transformed into a conjunction of two existential properties
\begin{alltt}
... 
(exists I:Point, incid\_lp I lAB \&\& incid\_lp I lCD) ->
        (exists J:Point,  \textbf{(incid\_lp J lAC \&\& incid\_lp J lBD)}) \ensuremath{/\backslash}
        (exists K:Point, \textbf{(incid\_lp K lAD \&\& incid\_lp K lBC)}). 
      \end{alltt}

      The tactic \texttt{wlog} was also extremely useful when dealing
      with spreads and trying to determine inside Coq which sets of
      lines are actual spreads.
\subsection{Producing Witnesses for Existential Proofs}
In the specification generator, we use a form of skolemisation to write functions
which compute the existential variable from the other arguments. For
instance, to achieve the proof of lemma \texttt{a3\_3}, we automatically
build a Coq function \texttt{f\_a3\_3} which, given three lines $l_1$,
$l_2$ and $l_3$ computes a line $l_4$ as well as its three
intersection points to lines  $l_1$,
$l_2$ and $l_3$.  
\begin{verbatim}
f_a3_3
     : Line -> Line -> Line -> Line * (Point * Point * Point)
\end{verbatim}

\subsection{Optimizing proofs}
We design some optimization techniques for generating and checking proof
  terms. We focus on the current goal, applying some sort of
  locality principle which means we try to prove a (sub-)goal the
  very first time we face it.
This means sequences of tactics such as
\begin{verbatim}
intros a; case a; intros H; 
  try (exact (degen_bool )_ H).
solve_goal.
\end{verbatim}
must be replaced by more efficient sequences like
\begin{verbatim}
intros a; case a; intros H; 
  solve [(exact (degen_bool )_ H |solve_goal].
\end{verbatim}
In this simplified example, we try to apply the tactic
\verb+(exact (degen_bool )_ H)+ for a subgoal and then we switch to the next
subgoal. Eventually we solve the remaining subgoals using the
\verb+solve_goal+ tactic.  The idea here is to solve the goal the
first time we encounter it. It is achieved by having several
possibilities of tactic applications to solve the goal (this
corresponds to the \verb+solve [t1|t2|t3]+ syntax.  The order of the
tactics \verb+t1+, \verb+t2+ and \verb+t3+ can be highly significant as well:
we should always call the tactic which is the most successful one on
such subgoals first.

As we face a huge number of cases, we need to design extremely
efficient prototype tactics on some specific subgoals and apply them automatically to all the subgoals
at stake.  Fine tuning the tactics rapidly is the key to making the
proofs faster to complete.

Finally, Coq provides some sort of task parallelism in the form of the
\texttt{par} tactical. It was very useful to deal with all the
sub-goals of a proof, once we figure out how to prove the first
one. The generic tactic proving the first goal, say \texttt{mytactic}
can be easily applied to all sub-goals in parallel (in some cases, we
have 35x35=1225 or more goals to deal with) by simply writing
\texttt{par: mytactic}.  

\section{Conclusion}\label{conclusion}
In this work, we manage to formalize in Coq the concepts of spreads and
packings. In the context of PG(3,2), we build automatically all the spreads and packings. We can easily verify that these generated sets of lines
(resp. spreads) are actual spreads (resp. packings). However it
remains challenging to verify that they are the only ones.  For spreads
of PG(3,2), we face a regression issue with Coq\footnote{
  \url{https://github.com/coq/coq/issues/13834}} which prevents our
proof from being accepted at the Qed step in the current version of Coq.

During this study, we faced case analysis with a huge
number of cases as well as debugging proof script with thousands of
sub-goals. We propose some proof engineering techniques to make
Coq process the files more easily, e.g . by directly providing witnesses for
example or by pruning the proof tree by using a \textsl{without loss of
  generality} principle.  

So far, we only address properties and transformations which remain in
the same (projective) space. We are currently working on generating specifications
of projective spaces automatically in order to easily have a formal
description of two different projective spaces and thus to be able to
formally describe constructions as the Bruck-Bose construction which
allows to build translation planes from projective planes \cite{BRUCK196485}.  


%
%
\bibliographystyle{eptcs}
\bibliography{generic}

\end{document}